\documentclass[paper]{JHEP3}

\usepackage{epsfig}
\usepackage{amssymb}
\usepackage{feynarts}
\def\URLtilde{\lower0.2em\hbox{$\tilde{\phantom{a}}$}}

\def\beq{\begin{equation}}
\def\eeq{\end{equation}}

\def\beqn{\begin{eqnarray}}
\def\eeqn{\end{eqnarray}}

\def\s#1{{\small#1}}
\def\HW{\s{HERWIG}}
\def\mhat{\widehat{m}}
\def\red{
}
\def\black{
}
\def\mycomm#1{\hfill\break\strut\kern-3em{\red\tt ====> #1\black}\hfill\break}

\newcommand{\ud}{\mathrm{d}}
\newcommand{\up}{\mathbf{p}}

\def\mqf{m^2_{qf}}

\preprint{Cavendish--HEP--05/11}
\title{Distinguishing Spins in Supersymmetric and Universal Extra Dimension
Models at the Large Hadron Collider%
\footnote{Work supported in part by the UK Particle Physics and
Astronomy Research Council.}}
\author{Jennifer M.\ Smillie$^1$ and Bryan R.\ Webber$^2$\\
  Cavendish Laboratory, University of Cambridge\\
  Madingley Road, Cambridge CB3 0HE, U.K.\\
  $^1$E-mail: \email{smillie@hep.phy.cam.ac.uk}\\
  $^2$E-mail: \email{webber@hep.phy.cam.ac.uk}}
\abstract{An interesting alternative to supersymmetry (SUSY) for extending
physics beyond the Standard Model is a model with universal extra
dimensions (UED), in which the SUSY superpartners are replaced
by Kaluza-Klein excitations of the Standard Model particles. If new
particles are discovered at the LHC, even if their mass spectrum
favours SUSY or UED, it will be vital to distinguish between their
spin assignments in the two models as far as possible.  We extend
the method proposed by Barr~\cite{Barr:2004ze} to the UED case and
investigate the angular and charge asymmetries of decay distributions
for sample mass spectra of both SUSY and UED types. For hierarchical
(`SUSY-type') mass spectra there is a good chance of distinguishing the
spin structures of the two models.  However, a mass spectrum of
the quasi-degenerate type expected in UED would make it
difficult to observe spin correlations.
  }

\begin{document}

%%%%%%%%%%%%%%%%%%%%%%%%%%%%%%%%%%%%%%%%%%%%%%%%%%%%%%%%%%%%%%%%%%%%%%
%%%%%%%%%%%%%%%%%%%%%%%%%% introduction %%%%%%%%%%%%%%%%%%%%%%%%%%%%%%
%%%%%%%%%%%%%%%%%%%%%%%%%%%%%%%%%%%%%%%%%%%%%%%%%%%%%%%%%%%%%%%%%%%%%%

\section{Introduction}\label{sec:intro}

The search for physics beyond the Standard Model (SM) will be a principal
objective of the main experiments at the CERN Large Hadron Collider
(LHC).  The leading contender amongst theories of new physics is
undoubtedly supersymmetry (SUSY), in which the SM particles
have massive superpartners that differ from them by a half-unit of spin.
The partners of strongly-interacting partons (the spartons) should be
copiously produced once their production thresholds are passed. In
R-parity conserving SUSY, the spartons typically decay into partons
and electroweak sparticles, which themselves decay sequentially through
the emission of SM particles to the stable lightest
supersymmetric particle (LSP). The LSP is usually the lightest
neutralino $\widetilde\chi^0_1$, which escapes from the detector
unobserved.  Thus the expected signature of SUSY is the copious
production of high-energy jets and leptons plus large missing energy.

If the above signature of SUSY is observed at the LHC, it will
be essential to confirm as far as possible that the produced objects
are indeed superpartners and not some other manifestation of new
physics.  Methods are being developed for exploring the mass spectrum
and decay systematics of candidate superparticles.\footnote{See for
example refs.~\cite{Allanach:2000kt,Lester:2001zx}.} The key issue then
will be whether their spins fit the pattern expected for SUSY.

In this connection, a very interesting alternative hypothesis is that
the new objects are indeed partners of the SM particles but
with {\em the same spins}. This scenario is realized in the so-called
universal extra dimension (UED) model, of the type proposed in
refs.~\cite{Appelquist:2000nn}.  It is remarkable that UED and SUSY models
could look very similar in collider experiments. In the former, the
Kaluza-Klein (KK) excitations of the SM particles can carry a KK-parity
analogous to the R-parity of the latter, implying pair production of
the first KK-excitations and a stable lightest KK-particle (LKP),
usually the first KK-excitation of the photon.  If the energy of
the LHC is insufficient to produce higher KK-excitations, then
each SM particle would be found to have a single KK-partner with
the same spin but a higher mass related to the size of the extra
dimension. 

It is not our intention here to advocate or criticise UED models as
a viable alternative to SUSY, but simply to discuss the extent to which
they could be distinguished from SUSY models with the same mass spectra
and decay systematics.  Admittedly certain mass spectra would already be
suggestive of one model or the other: for example, the spectrum
of KK-excitations is quasi-degenerate if the extra dimension is simply
compactified on an $S_1/Z_2$ orbifold~\cite{Cheng:2002iz}.  The
degeneracy is broken only by zero-mode (SM) masses, volume-suppressed
boundary terms and loop corrections with a relatively low cutoff.
In SUSY models one usually assumes that soft SUSY-breaking terms
satisfy universal boundary conditions at a much higher scale, so
that such degeneracy at the weak scale would be unnatural.
However, the number of
unknowns and arbitrary assumptions in each case is so great that
the exclusion of either class of model in favour of the other would
only be truly convincing if their spin structures could be distinguished.

In the present paper we assume that a particular chain of decays that
is common in both SUSY and UED models, starting from a squark or a
KK-excited quark respectively, has been identified and that all
the masses of the new particles involved in it are known.  We then study
the extent to which decay correlations, manifest in the invariant mass
distributions of combinations of observable decay products, would
enable one to distinguish between the SUSY and UED spin assignments
of the new particles.  Our work is thus an extension of that described
in ref.~\cite{Barr:2004ze} (see also~\cite{Goto:2004cp}),
where the SUSY decay correlations were
compared with uncorrelated phase space.\footnote{Distinguishing between
SUSY and UED at future $e^+e^-$ colliders has been studied in
refs.~\cite{Bhattacharyya:2005vm,Battaglia:2005zf}.}

In the following section we present the decay chains to be considered, and in
section~\ref{sec:susyspin} we recall the SUSY correlations investigated earlier.  In
section~\ref{sec:uedspin} we present our new results on the corresponding UED decay
correlations.  We derive simple analytical formulae for the correlation coefficients in
terms of the masses in the decay chain, which should be of general use whatever the mass
spectrum might turn out to be.  We show graphical results for two specific mass scenarios,
one considered more probable in SUSY and the other in UED.  In both mass scenarios we
compare the correlations predicted by the SUSY and UED spin assignments.

As was emphasised by Barr~\cite{Barr:2004ze}, the observability
of interesting correlations depends crucially on the fact that
the LHC is a proton-proton collider, so that squarks/KK-quarks are
produced somewhat more copiously than their antiparticles.  To
quantify this effect, we need to know the direct and indirect
production cross sections of KK-quarks and KK-antiquarks. We
have therefore computed the lowest-order two-parton to two-KK-parton
matrix elements, which are expected to dominate the production of these
particles.  Our results, which differ somewhat from those presented in
ref.~\cite{Macesanu:2002db},\footnote{An erratum to ref.~\cite{Macesanu:2002db}
is in preparation (C.~Macesanu, private communication).} are discussed
in section~\ref{sec:prod} and listed in appendix~\ref{app:prod}.

Using our results on the UED production matrix elements and decay
correlations, together with the decay branching ratios suggested in
ref.~\cite{Cheng:2002iz}, we have included a full simulation of
the relevant UED processes in the \HW\ Monte Carlo event
generator~\cite{Corcella:2000bw,Corcella:2002jc}. Since the corresponding
SUSY processes, with full spin correlations, are already a well-established
feature of \HW~\cite{Richardson:2001df,Moretti:2002eu}, we are able in
section~\ref{sec:exp} to present first  detector-level results on distinguishing
UED and SUSY spin correlations at the LHC.  Our results and conclusions
are summarized in section~\ref{sec:conc}.

%%%%%%%%%%%%%%%%%%%%%%%%%%%%%%%%%%%%%%%%%%%%%%%%%%%%%%%%%%%%%%%%%%%%%%
%%%%%%%%%%%%%%%%%%%%%%%%%%%% sections %%%%%%%%%%%%%%%%%%%%%%%%%%%%%%%%
%%%%%%%%%%%%%%%%%%%%%%%%%%%%%%%%%%%%%%%%%%%%%%%%%%%%%%%%%%%%%%%%%%%%%%

\section{Decay chains in SUSY and UED}\label{sec:dec}

\FIGURE{
\unitlength=0.5bp%
\put(2,1){(a)}
%\begin{feynartspicture}(432,504)(1,1)
\begin{feynartspicture}(432,380)(1,1)

\FADiagram{}
\FAProp(0.,10.)(5.5,13.5)(0.,){/ScalarDash}{0}
\FALabel(2.39784,12.5777)[br]{$\widetilde q_L$}
\FAProp(5.5,19.5)(5.5,13.5)(0.,){/Straight}{-1}
\FALabel(4.43,16.5)[r]{$q_L$}
\FAProp(13.,17.5)(11.,13.)(0.,){/Straight}{1}
\FALabel(11.0636,15.3995)[br]{$l^{\rm near}$}
\FAProp(20.,12.5)(14.5,11.)(0.,){/Straight}{-1}
\FALabel(16.8422,12.7654)[b]{$l^{\rm far}$}
\FAProp(16.,5.5)(14.5,11.)(0.,){/Straight}{-1}
\FALabel(16.2654,8.65783)[l]{$\widetilde\chi^0_1$}
\FAProp(5.5,13.5)(11.,13.)(0.,){/Straight}{1}
\FALabel(8.10967,12.1864)[t]{$\widetilde\chi^0_2$}
\FAProp(11.,13.)(14.5,11.)(0.,){/ScalarDash}{0}
\FALabel(12.461,11.1342)[tr]{$\tilde l$}
\FAVert(5.5,13.5){0}
\FAVert(11.,13.){0}
\FAVert(14.5,11.){0}
\end{feynartspicture}

\put(2,1){(b)}
\begin{feynartspicture}(432,380)(1,1)

\FADiagram{}
\FAProp(0.,10.)(5.5,13.5)(0.,){/Straight}{1}
\FALabel(2.39784,12.5777)[br]{$q^*_L$}
\FAProp(5.5,19.5)(5.5,13.5)(0.,){/Straight}{-1}
\FALabel(4.43,16.5)[r]{$q_L$}
\FAProp(13.,17.5)(11.,13.)(0.,){/Straight}{1}
\FALabel(11.0636,15.3995)[br]{$l^{\rm near}$}
\FAProp(20.,12.5)(14.5,11.)(0.,){/Straight}{-1}
\FALabel(16.8422,12.7654)[b]{$l^{\rm far}$}
\FAProp(16.,5.5)(14.5,11.)(0.,){/Sine}{0}
\FALabel(16.2654,8.65783)[l]{$\gamma^*$}
\FAProp(5.5,13.5)(11.,13.)(0.,){/Sine}{0}
\FALabel(8.10967,12.1864)[t]{$Z^*$}
\FAProp(11.,13.)(14.5,11.)(0.,){/Straight}{1}
\FALabel(12.461,11.1342)[tr]{$l^*$}
\FAVert(5.5,13.5){0}
\FAVert(11.,13.){0}
\FAVert(14.5,11.){0}
\end{feynartspicture}
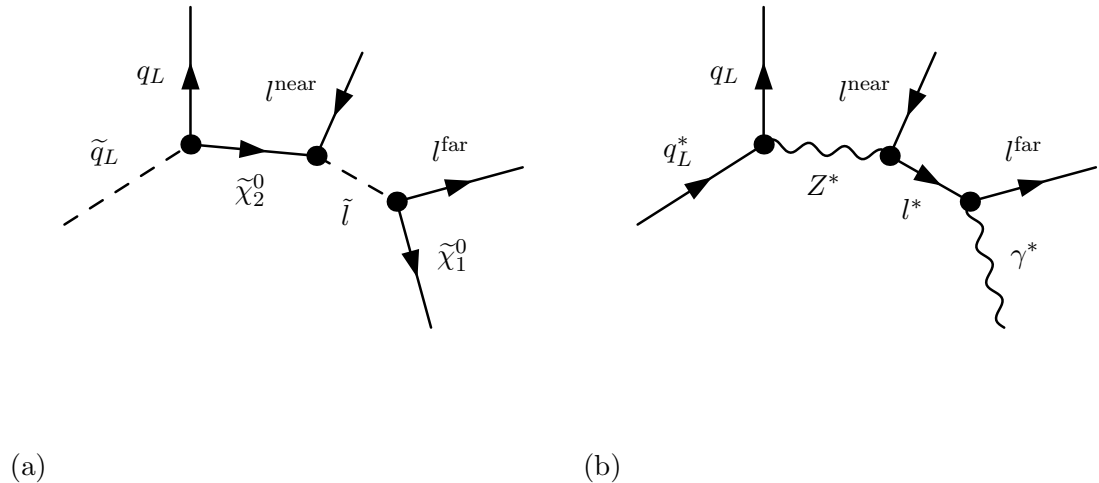
\caption{(a) SUSY and (b) UED decay chains considered here.
\label{fig:decay_chains}}
}

The SUSY decay chain that we shall consider, which is the same
as that studied in ref.~\cite{Barr:2004ze}, is shown in
figure~\ref{fig:decay_chains}, together with the corresponding UED process.
In both cases the visible decay products are a quark jet and a pair of
opposite-sign same-flavour (OSSF) leptons with the same chirality.
We suppose that the new particle masses have been measured, either by an
edge analysis along the lines of refs.~\cite{Allanach:2000kt,Lester:2001zx}
or some other means, and it remains to decide whether the decay angular
distributions agree better with the SUSY or UED spin assignments.

The angular distributions depend on whether or not the chirality of the
slepton/KK-lepton is the same at that of the decaying
squark/KK-quark.\footnote{We should emphasise that we use
the term `chirality' loosely
here, since neither the sparticles nor the KK-excitations concerned
have definite handedness: what we mean is that they couple to SM particles
of that chirality.}
For definiteness, we assume that the latter is left-handed, which is
preferred in both of the models under consideration. We can then
characterise the process by the chirality and charge of the ``near'' lepton,
defined as shown in figure~\ref{fig:decay_chains}.  Of course, we cannot
distinguish experimentally between the ``near'' and ``far'' leptons,
and so their contributions to any distribution will eventually have to
be combined.  However, in principle (in the zero-width approximation that
we use) the processes with opposite  ``near'' and ``far'' charge assignments
are distinct.  There are then two fundamental processes with different
decay correlations, which (as in ref.~\cite{Barr:2004ze}) we label 1 and 2:
\begin{itemize}
\item Process 1: $\{ q,l^{\rm near},l^{\rm far}\} = 
    \{ q_L, l^-_L,l^+_L \}$ or $\{\bar q_L,l^+_L,l^-_L\}$
or $\{ q_L, l^+_R,l^-_R \}$ or $\{\bar q_L,l^-_R,l^+_R\}$;
\item Process 2: $\{ q,l^{\rm near},l^{\rm far}\} = 
    \{ q_L, l^+_L,l^-_L \}$ or $\{\bar q_L,l^-_L,l^+_L\}$
or $\{ q_L, l^-_R,l^+_R \}$ or $\{\bar q_L,l^+_R,l^-_R\}$.
\end{itemize}

\section{Spin correlations in SUSY}\label{sec:susyspin}

We first recapitulate from ref.~\cite{Barr:2004ze} the angular distributions
that are expected in the SUSY decay chain \ref{fig:decay_chains}(a).
The $\widetilde\chi^0_2$ has spin one-half and its helicity is the same as that
of the quark, since the squark is a scalar.  Therefore a near lepton with
the same helicity as the quark (process 1) will be emitted preferentially
at large values of the angle $\theta^*$ between its direction and that of
the quark in the $\widetilde\chi^0_2$ rest frame, with angular distribution
(neglecting all SM particle masses)
\beq
\frac{\ud P^{\rm SUSY}_1}{\ud\cos\theta^*} = \frac 12(1-\cos\theta^*)\;.
\eeq
A near lepton with helicity opposite to the quark (process 2), on the other hand,
will have angular distribution
\beq
\frac{\ud P^{\rm SUSY}_2}{\ud\cos\theta^*} = \frac 12(1+\cos\theta^*)\;.
\eeq
In terms of the $q l^{\rm near}$ invariant mass,
\beq
(m_{ql}^{\rm near})^2=2 \vert \up_l \vert \vert \up_q \vert(1-\cos \theta^*) =
\frac 12(m_{ql}^{\rm near})^2_{\rm max} (1-\cos \theta^*)\;,
\eeq
defining the rescaled invariant mass variable to be 
\beq\label{eq:mhat}
\mhat=~m_{ql}^{\rm near}/(m_{ql}^{\rm near})_{\rm max}=\sin(\theta^*/2)
\eeq
we therefore have
\beq\label{eq:SUSY_P1}
\frac{\ud P^{\rm SUSY}_1}{\ud \mhat} = 4 \mhat^3
\eeq
and
\beq\label{eq:SUSY_P2}
\frac{\ud P^{\rm SUSY}_2}{\ud \mhat} = 4 \mhat(1-\mhat^2)\;.
\eeq

The slepton produced in the decay of the $\widetilde\chi^0_2$ is a scalar
particle, and so its decay is isotropic in its rest frame, and the near and
far lepton directions are uncorrelated in that frame. There is nevertheless a
weak correlation between the directions of the far lepton and the quark,
due to the relative velocity of the slepton and the $\widetilde\chi^0_2$,
which we discuss in more detail in appendix~\ref{app:qlfar}.

\section{Spin correlations in UED}\label{sec:uedspin}

\subsection{Correlations in $q^*$ and $Z^*$ decays}

In the UED decay chain, the primary object has spin one-half and it
decays to a quark and a vector boson, $q^*\to qZ^*$.
% This is analogous to the decay of the top quark, $t\to bW$. 
The vector boson is either longitudinally
or transversely polarised, with relative probabilities $m_{q^*}^2$ and
$2m_{Z^*}^2$, respectively. By angular momentum conservation, when the
polarisation is transverse the helicity of the $Z^*$, and of the near
lepton, must be the same as that of the quark.  The distribution of
the angle $\theta^*$ between the directions of the near lepton
and the quark in the $Z^*$ rest frame is therefore of the form
\beq
\frac{\ud P^{\rm UED}_{1,2}}{\ud\cos\theta^*} =\frac{1}{1+2x}\left(
\frac{\ud P_0}{\ud\cos\theta^*}+2x\frac{\ud P_{\mp}}{\ud\cos\theta^*}
\right)
\eeq
where $x=m_{Z^*}^2/m_{q^*}^2$ and $P_\lambda$ on the right-hand side
represents the distribution for $Z^*$ helicity $\lambda$.  

In the rest frame of the $Z^*$, the decay of
the longitudinally polarised state is forward-backward symmetric, with
angular distribution
\beq
\frac{\ud P_0}{\ud\cos\theta^*} = \frac{3}{2(2+y)}(\sin^2\theta^*
+y\cos^2\theta^*)
\eeq
where  $y=m_{l^*}^2/m_{Z^*}^2$.  The transverse decay distribution for
$\pm$ve helicity is
\beq
\frac{\ud P_\pm}{\ud\cos\theta^*} = \frac{3}{4(2+y)}[(1\pm\cos\theta^*)^2
+y\sin^2\theta^*]\;.
\eeq
The angular distributions for the two fundamental processes are therefore given by
\beq\label{eq:UEDstar}
\frac{\ud P^{\rm UED}_{1,2}}{\ud\cos\theta^*} = 
\frac{3}{2(1+2x)(2+y)}\left[1+x+xy\mp 2x\cos\theta^*
-(1-x)(1-y)\cos^2\theta^*\right]\;.
\eeq

\subsection{Quark + near lepton mass distribution}

Using eq.~(\ref{eq:UEDstar}), the distribution of the rescaled
$q l^{\rm near}$ invariant mass (\ref{eq:mhat}) is given in process 1 by
\beq\label{eq:UED_P1}
\frac{\ud P^{\rm UED}_1}{\ud \mhat} =\frac{6\mhat}
{(1+2x)(2+y)}\left[y+4(1-y+xy)\mhat^2-4(1-x)(1-y)\mhat^4\right]
\eeq
and in process 2 by
\beq\label{eq:UED_P2}
\frac{\ud P^{\rm UED}_2}{\ud \mhat} =\frac{6\mhat}
{(1+2x)(2+y)}\left[4x+y+4(1-2x-y+xy)\mhat^2-4(1-x)(1-y)\mhat^4\right]\;.
\eeq

Notice that the $\mhat$ distributions for the two UED processes become
identical as $x\to 0$, i.e. for $m_{Z^*}\ll m_{q^*}$, since in that limit
the $Z^*$ polarisation is purely longitudinal.  On the other hand when
$x\to 1$ and $y\to 0$ ($m_{l^*}\ll m_{Z^*}\sim m_{q^*}$) the
UED distributions become indistinguishable from those of the
corresponding SUSY processes, eqs.~(\ref{eq:SUSY_P1},\ref{eq:SUSY_P2}).

The features of the $ql^{\rm near}$ mass distributions can be illustrated
by considering their forms for typical UED and SUSY patterns of new particle
masses. A possible UED mass spectrum (from \cite{Cheng:2002iz}) is given in
table~\ref{tab:UEDmasses} for inverse radius $R^{-1}=500$GeV, cut-off $\Lambda$
such that $\Lambda R=20$ and $m_h=120$ GeV.  This model also assumes vanishing
boundary terms at the cut-off scale $\Lambda$, and a vanishing boundary mass
term for the Higgs mass, $\overline{m}_h^2$.  The lightest four left KK-quarks
are degenerate in mass and are labelled here collectively as $q_L^*$.
Similarly the right and left KK-electrons and KK-muons are degenerate in mass
and are labelled here as $l_R^*$ and $l_L^*$ respectively.  This spectrum
illustrates the feature of UED that the new particles have masses which are
much closer to each other (usually within $100-200$ GeV) than in a
SUSY spectrum based on high-scale universality.
\TABLE{
\begin{tabular}{|c|c|c|c|c|}
\hline $\gamma^*$ & $Z^*$ & $q_L^*$ & $l_R^*$ & $l_L^*$ \\ \hline
501&536&598&505&515\\ \hline
\end{tabular}
\caption{UED masses in GeV, for $R^{-1}=500$GeV, $\Lambda R=20$, $m_h=120$GeV,
$\overline{m}_h^2=0$ and vanishing boundary terms at cut-off scale $\Lambda$.
\label{tab:UEDmasses}}
}

In the UED model we have eqs.~(\ref{eq:UED_P1}) and (\ref{eq:UED_P2})
with $x=m_{Z^*}^2/m_{q^*}^2=0.803$; the $Z^*$ decays preferentially
to a left-handed excited lepton and so we use $y=m_{l_L^*}^2/m_{Z^*}^2
=0.923$, which yields
\beqn 
\frac{\ud P^{\rm UED}_1}{\ud \mhat}&=&
0.727 \mhat+2.577 \mhat^3-0.047 \mhat^5\;,\nonumber\\ 
\frac{\ud P^{\rm UED}_2}{\ud \mhat}&=& 3.257
\mhat-2.483 \mhat^3-0.047 \mhat^5\;. 
\eeqn 
These should be compared with the corresponding SUSY expressions
(\ref{eq:SUSY_P1}) and (\ref{eq:SUSY_P2}), which are independent of
the particle masses.

\FIGURE{ \put(20,0){(a)} \epsfig{figure=ued1.ps,height=0.25\textheight}
  \put(20,0){(b)}\epsfig{figure=ued2.ps,height=0.25\textheight}
\caption{UED and SUSY distributions for (a) Process 1 and (b) Process 2 with respect to the
rescaled $q l^{\rm near}$ invariant mass, for the UED mass spectrum in
table~\ref{tab:UEDmasses}. Dotted: phase space. Dashed: SUSY. Solid/red: UED.
\label{fig:UED-qlnear}}
}

The UED and SUSY angular distributions are plotted against each other for processes 1 and
2 in figures \ref{fig:UED-qlnear}(a) and \ref{fig:UED-qlnear}(b) respectively.  Since
$x=m_{Z^*}^2/m_{q^*}^2$ is large in the typical UED scenario, and the effect of
$y=m_{l^*}^2/m_{Z^*}^2$ is weak at large $x$, the UED and SUSY distributions are similar.
Therefore it will be difficult to verify the UED spin assignments if the spectrum is
quasi-degenerate like that in table~\ref{tab:UEDmasses}.

\TABLE{
\begin{tabular}{|c|c|c|c|c|}
\hline $\widetilde{\chi}_1^0$ & $\widetilde{\chi}_2^0$ & $\widetilde{u}_L$ & $\widetilde{e}_R$
& $\widetilde{e}_L$ \\ \hline
96&177&537&143&202\\ \hline
\end{tabular}
\caption{SUSY masses in GeV, for SPS point 1a.
\label{tab:SUSYmasses}}
}

The SUSY mass spectrum, on the other hand, does not naturally have the
same near-degeneracy of neutralinos and squarks, and therefore the UED
and SUSY  angular distributions are more distinct.  For illustration,
we consider the MSSM Snowmass point SPS 1a~\cite{Allanach:2002nj},
which has the mass spectrum shown in table~\ref{tab:SUSYmasses}.  The decay 
$\widetilde{\chi}_2^0\to l\tilde l_R$ is preferred and therefore we use
$x=m_{\widetilde{\chi}_2^0}^2/m_{\widetilde{u}_L}^2=0.109$ and
$y=m_{\widetilde e_R}^2/m_{\widetilde{\chi}_2^0}^2=0.653$ for the
comparative UED distributions, giving
\beqn 
\frac{\ud P^{\rm UED}_1}{\ud \mhat}&=&
1.213 \mhat+3.108 \mhat^3-2.301 \mhat^5\;,\nonumber\\ 
\frac{\ud P^{\rm UED}_2}{\ud \mhat}&=& 2.020
\mhat+1.493 \mhat^3-2.301 \mhat^5\;. 
\eeqn

\FIGURE{\put(20,0){(a)} \epsfig{figure=sps1.ps,height=0.25\textheight}
  \put(20,0){(b)}\epsfig{figure=sps2.ps,height=0.25\textheight}
\caption{UED and SUSY distributions for (a) Process 1 and (b) Process 2 with respect to
the rescaled $q l^{\rm near}$ invariant mass, for the SUSY mass spectrum in
table~\ref{tab:SUSYmasses}. Dotted: phase space. Dashed: SUSY. Solid/red: UED.
\label{fig:SUSY-q lnear}}
}

The resulting mass distributions are compared in
fig.~\ref{fig:SUSY-q lnear}. Owing to the small value of $x$, the
UED predictions for the two processes are similar to each other, and
different from the SUSY predictions.  This gives some grounds for
optimism that, if the spectrum is hierarchical, like that in
table~\ref{tab:SUSYmasses}, then the SUSY spin assignments can be
confirmed or ruled out in comparison with the UED assignments.

\subsection{Correlations in $l^*$ decay}
In the SUSY decay chain (figure 1a) , the slepton $\tilde l$ is spinless and therefore
it decays isotropically in its rest frame.   In the UED case (figure 1b), the spin of the
KK lepton $l^*$ induces non-trivial correlations. Up to an overall constant, the full
matrix elements for UED processes 1 and 2, as defined in section~\ref{sec:dec},
take the form
\beq\label{eq:uedspin}
 |\overline{\mathcal{M}}|^2 \propto 2z(1-z)W_{l^*}+(1-2z)W_f
 \eeq
 where $z=m_{\gamma^*}^2/m_{l^*}^2$, $f$ represents the far lepton and, for $l=l^*$ or $f$, 
  \beqn
 W_l&=& (1-x)(2p_{Z^*}\cdot p_n\,p_{Z^*}\cdot p_l+m_{Z^*}^2\,p_n\cdot p_l)
 -4x(p_n\cdot p_q\,p_{Z^*}\cdot p_l+p_n\cdot p_{Z^*}\,p_q\cdot p_l) +\nonumber\\
  &+& \left\{\begin{array}{c}
  8x^2\,p_n\cdot p_q\,p_{q^*}\cdot p_l\;\;\mbox{for process 1} \\
  8x^2\,p_n\cdot p_{q^*}\,p_q\cdot p_l\;\;\mbox{for process 2}\end{array}\right.
\eeqn
 where, as before, $x=m_{Z^*}^2/m_{q^*}^2$, and $n$ represents the near lepton.

To specify the $l^*$ decay distribution, we define $\theta$ as the angle between
the near and far leptons in the $l^*$ rest frame, and $\phi$ as the angle between
the $q l^{\rm near}$ and dilepton planes, in the same frame.   Then we find
\beqn
\frac{\ud^3 P^{\rm UED}_{1,2}}{\ud\cos\theta^*\,\ud\cos\theta\,\ud\phi} &=& 
\frac{3}{4\pi(1+2x)(2+y)}\Bigl[1+x+xy\mp 2x\cos\theta^*
-(1-x)(1-y)\cos^2\theta^*\nonumber\\
&&-\frac{1-2z}{1+2z}\Bigl\{\left[1+x-xy\mp 2x\cos\theta^*
-(1-x)(1+y)\cos^2\theta^*\right]\cos\theta\nonumber\\
&&-2\left[(1-x)\cos\theta^*\pm x\right]
\sqrt{y}\sin\theta^*\sin\theta\cos\phi\Bigr\}\Bigr]\;.
\eeqn

\subsection{Dilepton mass distribution}
The dilepton mass $m_{ll}$ is simply related to the $l^*$ decay angle $\theta$:
\beq
m_{ll}^2 = \frac 14 x^2(1-y)(1-z)(1-\cos\theta)\,m_{q^*}^2
\eeq
and so, defining 
\beq\label{eq:mhatll}
\mhat_{ll}= m_{ll}/(m_{ll})_{\rm max}=\sin(\theta/2)
\eeq
we have
\beq\label{eq:UED_ll}
\frac{\ud P^{\rm UED}_{1,2}}{\ud \mhat_{ll}} =\frac{4\mhat_{ll}}
{(2+y)(1+2z)}[y+4z+(2-y)(1-2z)\mhat_{ll}^2]\;.
\eeq

The dilepton mass distribution is potentially a good indicator of UED spin correlations,
because it is the same for processes 1 and 2 and relatively easy to measure.  We see from
eq.~(\ref{eq:UED_ll}) that the deviation from the linear mass spectrum of phase space
or SUSY is greatest when
$y$ and $z$ are small, i.e., when $m_{\gamma^*}\ll m_{l^*}\ll m_{Z^*}$.  On the
other hand, the spin correlation vanishes when $z=\frac 12$, that is, when $m_{\gamma^*}
=m_{l^*}/\sqrt{2}$, which is close to the situation at SPS point 1a.  The typical UED
scenario with quasi-degenerate masses also yields a small effect, with a mass
spectrum proportional to $\mhat_{ll}(1-\mhat_{ll}^2/5)$.  Therefore a dilepton
mass distribution differing significantly from phase space or SUSY would only be
manifest in a UED model substantially different from those considered so far.

\FIGURE{ \put(20,0){(a)} \epsfig{figure=ued8.ps,height=0.25\textheight}
  \put(20,0){(b)}\epsfig{figure=sps8.ps,height=0.25\textheight}
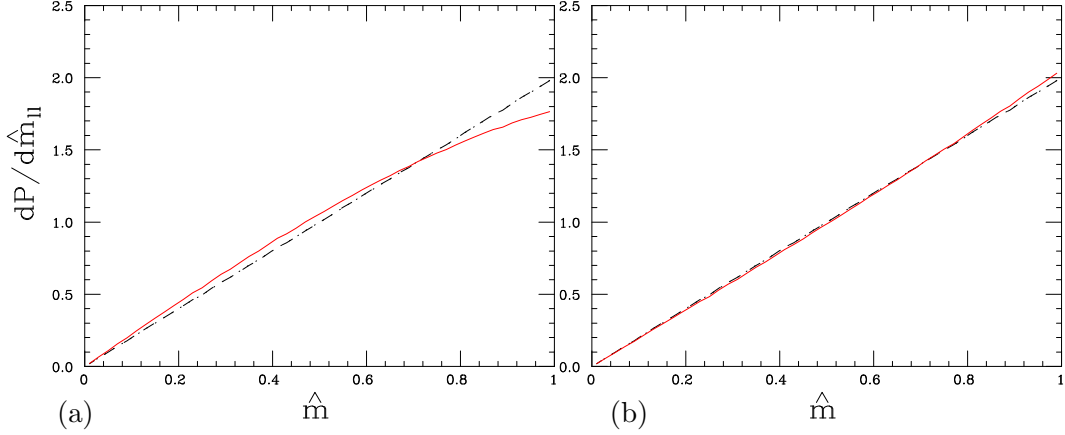
\caption{UED and SUSY distributions with respect to the rescaled dilepton
invariant mass, for (a) the UED and (b) the SUSY mass spectrum given above.
Dotted: phase space. Dashed: SUSY. Solid/red: UED.
\label{fig:lnlf}}
}

\subsection{Quark + far lepton mass distribution}\label{sec:qlfar}

The quark + far lepton mass is a function of all the decay angles,
\beqn
(m_{ql}^{\rm far})^2 &=& \frac 14 (1-x)(1-z)\Bigl[(1+y)(1-\cos\theta^*\cos\theta)+\nonumber\\
&&+(1-y)(\cos\theta^*-\cos\theta)-2\sqrt{y}\sin\theta^*\sin\theta\cos\phi\Bigr]\,m_{q^*}^2\;.
\eeqn
The maximum value occurs for $\theta^*=0$ and $\theta=\pi$, when
\beq
(m_{ql}^{\rm far})^2_{\rm max} = (1-x)(1-z)\,m_{q^*}^2\;,
\eeq
and therefore the rescaled quark + far lepton mass is given by
\beqn\label{eq:mqlfar}
\mhat_{qf} \equiv m_{ql}^{\rm far}/(m_{ql}^{\rm far})_{\rm max}
&=& \frac 12\Bigl[(1+y)(1-\cos\theta^*\cos\theta)+\nonumber\\
&&+(1-y)(\cos\theta^*-\cos\theta)-2\sqrt{y}\sin\theta^*\sin\theta\cos\phi\Bigr]^{\frac 12}\;.
\eeqn

The phase space for this quantity increases linearly up to the point $\mhat_{qf}
=\sqrt y$, then decreases logarithmically to zero.  In the region $\mhat_{qf}<\sqrt y$, the
probability distribution has a polynomial form, as illustrated in figures~\ref{fig:UED-qlfar}
and \ref{fig:SPS-qlfar} for the UED and SUSY mass scenarios respectively.
Equations for the distribution are given in appendix~\ref{app:qlfar}.

\FIGURE{ \put(20,0){(a)} \epsfig{figure=ued3.ps,height=0.25\textheight}
  \put(20,0){(b)}\epsfig{figure=ued4.ps,height=0.25\textheight}
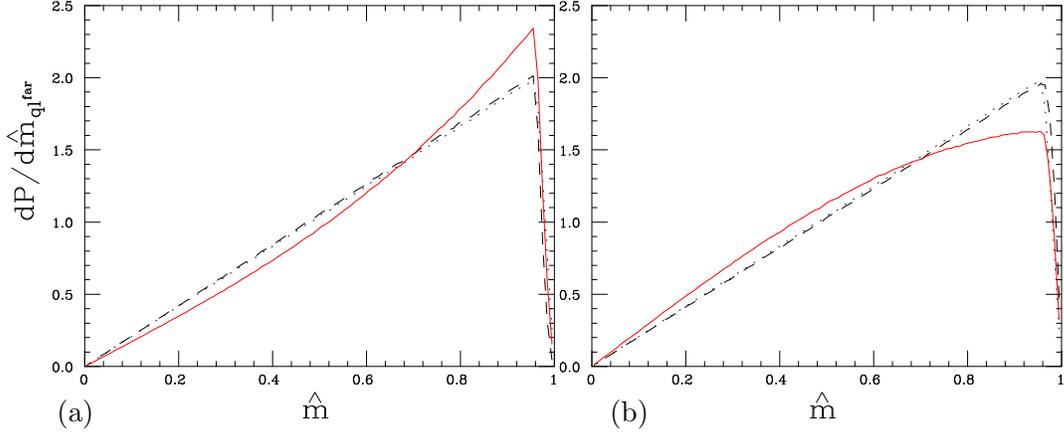
\caption{UED and SUSY distributions for (a) Process 1 and (b) Process 2 with respect to the
rescaled $ql^{\rm far}$ invariant mass, for the UED mass spectrum in
table~\ref{tab:UEDmasses}. Dotted: phase space. Dashed: SUSY. Solid/red: UED.
\label{fig:UED-qlfar}}
}
\FIGURE{ \put(20,0){(a)} \epsfig{figure=sps3.ps,height=0.25\textheight}
  \put(20,0){(b)}\epsfig{figure=sps4.ps,height=0.25\textheight}
\caption{UED and SUSY distributions for (a) Process 1 and (b) Process 2 with respect to the
rescsled $q l^{\rm far}$ invariant mass, for the SUSY mass spectrum in
table~\ref{tab:SUSYmasses}. Dotted: phase space. Dashed: SUSY. Solid/red: UED.
\label{fig:SPS-qlfar}}
}

We see that, as in the case of the dilepton mass distribution, the spin dependence of the
quark + far lepton distribution is weak for both the mass spectra considered here.

\subsection{Observable quark-lepton correlations}

To proceed further, we must face the fact that the $ql^{\rm near}$
and $ql^{\rm far}$ mass distributions are not experimentally observable. 
As pointed out
in ref.~\cite{Barr:2004ze}, the best that can be done in reality is
to measure the invariant masses of jet + lepton combinations.
Assuming that the jet and lepton are indeed decay products from
process 1 or 2, the $jl^\pm$ mass distribution for a given lepton
charge receives near-lepton contributions from the corresponding
process and the charge conjugate of the other process, plus far-lepton
contributions from the other process and the charge conjugate of the
same process. Concentrating on the UED scenario of a left-handed
KK-quark decaying to a left-handed KK-lepton, we have
\beq\label{eq:mjl+}
\frac{\ud P}{\ud m_{jl^+}}
= f_q\left(
\frac{\ud P_2}{\ud m_{ql}^{\rm near}}+ 
\frac{\ud P_1}{\ud m_{ql}^{\rm far}}\right)
+ f_{\bar q}\left(
\frac{\ud P_1}{\ud m_{ql}^{\rm near}}+ 
\frac{\ud P_2}{\ud m_{ql}^{\rm far}}\right)
\eeq
where $f_q$ and $f_{\bar q}$ are the quark and antiquark fractions
in the selected event sample.  Similarly
\beq\label{eq:mjl-}
\frac{\ud P}{\ud m_{jl^-}}
= f_q\left(
\frac{\ud P_1}{\ud m_{ql}^{\rm near}}+ 
\frac{\ud P_2}{\ud m_{ql}^{\rm far}}\right)
+ f_{\bar q}\left(
\frac{\ud P_2}{\ud m_{ql}^{\rm near}}+ 
\frac{\ud P_1}{\ud m_{ql}^{\rm far}}\right)\;.
\eeq
As will be discussed in section~\ref{sec:prod},
for both the UED and SUSY scenarios we find $f_q\simeq 0.7$, $f_{\bar q}\simeq 0.3$
at the LHC. The resulting observable jet+lepton mass distributions are then as depicted
in figure~\ref{fig:ued_mjl}, where again we have normalised to the maximum observable
mass.

\FIGURE{ \put(20,0){(a)} \epsfig{figure=ued_mjl+.ps,height=0.25\textheight}
  \put(20,0){(b)}\epsfig{figure=ued_mjl-.ps,height=0.25\textheight}
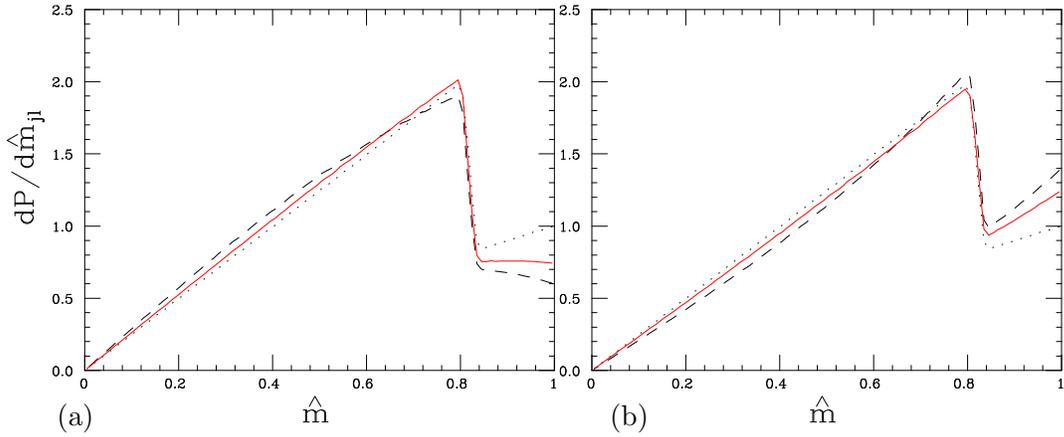
\caption{UED and SUSY rescaled mass distributions for (a) jet + $l^+$ (b) jet + $l^-$,
for the UED mass spectrum in table~\ref{tab:UEDmasses}. Dotted: phase space. Dashed:
SUSY. Solid/red: UED.
\label{fig:ued_mjl}}
}

Corresponding results for the SUSY mass scenario are shown in figure~\ref{fig:sps_mjl}.
Here the roles of processes 1 and 2 are interchanged in eqs.~(\ref{eq:mjl+}) and
(\ref{eq:mjl-}), since the decay now involves a right-handed slepton or KK-lepton.

\FIGURE{ \put(20,0){(a)} \epsfig{figure=sps_mjl+.ps,height=0.25\textheight}
  \put(20,0){(b)}\epsfig{figure=sps_mjl-.ps,height=0.25\textheight}
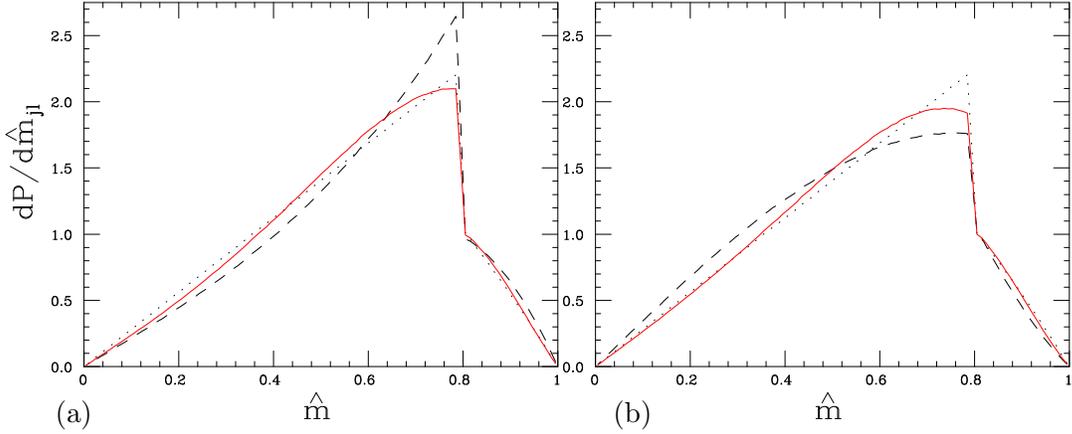
\caption{UED and SUSY rescaled mass distributions for (a) jet + $l^+$ (b) jet + $l^-$,
for the SUSY mass spectrum in table~\ref{tab:SUSYmasses}. Dotted: phase space. Dashed:
SUSY. Solid/red: UED.
\label{fig:sps_mjl}}
}

Figure~\ref{fig:asymm} shows the resulting charge asymmetry
\beq
A=\frac{\ud P/\ud m_{jl^+}-\ud P/\ud m_{jl^-}}
     {\ud P/\ud m_{jl^+}+\ud P/\ud m_{jl^-}}
\eeq
We see that the UED and SUSY charge asymmetries are similar in form, but the latter
is smaller by a factor of 2 -- 4.

\FIGURE{ \put(20,0){(a)} \epsfig{figure=ued7.ps,height=0.25\textheight}
  \put(20,0){(b)}\epsfig{figure=sps7.ps,height=0.25\textheight}
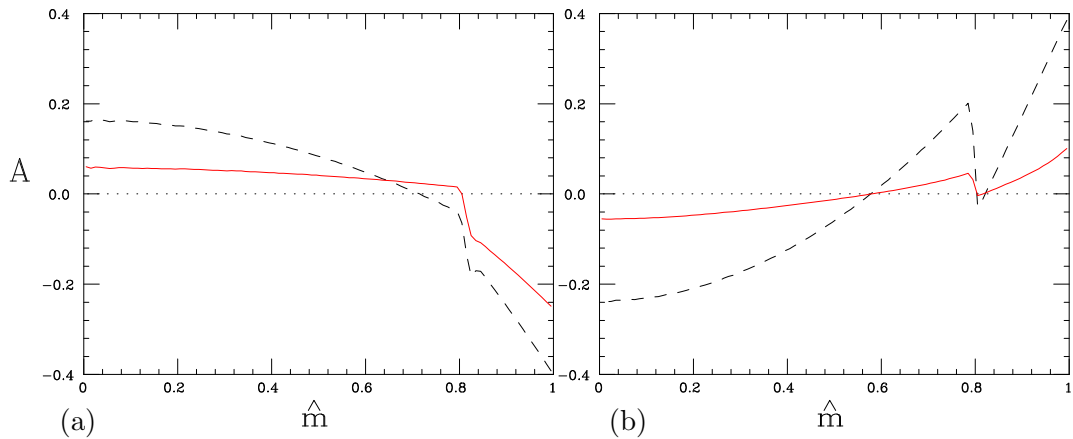
\caption{UED and SUSY charge asymmetries with respect to the jet + lepton rescaled invariant mass,
for (a) the UED and (b) the SUSY mass spectrum given above. Dotted: phase space. Dashed: SUSY.
Solid/red: UED.
\label{fig:asymm}}
}

\section{Production cross sections}\label{sec:prod}

As we saw in the previous section, the observability of a charge asymmetry sensitive to
UED spin correlations depends critically on the difference between the production rates
of KK-excited quarks and antiquarks, just as in SUSY it depends on the difference of
squark and antisquark rates~\cite{Barr:2004ze}. We have therefore computed the
relevant lowest-order two-to-two scattering subprocess matrix elements and used them,
together with the UED branching ratios suggested in ref.~\cite{Cheng:2002iz}, to
estimate the UED production cross sections and the quantities $f_q$ and $f_{\bar q}$
appearing in eqs.~(\ref{eq:mjl+}) and (\ref{eq:mjl-}).

Our expressions for the subprocess matrix elements are listed in appendix~\ref{app:prod}.
These results were obtained by including the Feynman rules for the effective
four-dimensional theory in \texttt{CompHEP} \cite{Pukhov:1999gg}. They differ
in several respects from those presented in ref.~\cite{Macesanu:2002db}.
Details of the discrepancies are given in the appendix.  Most importantly,
we find a larger overall normalization.

\TABLE{
\begin{tabular}{|c|c||c|c|c|c|}
\hline
Masses & Model & $\sigma_{\rm all}$ & $\sigma_{q^*}$ & $\sigma_{\bar q^*}$ & $f_q$ \\
\hline \hline
& & & & & \\
UED    & UED  & 249 & 158 & 83 & 0.66 \\
& & & & & \\
UED    & SUSY &  28 &  18 &  9 & 0.65 \\
& & & & & \\
SPS 1a & UED  & 480 & 230 & 102 & 0.69 \\
& & & & & \\
SPS 1a & SUSY &  55 &  26 & 11 & 0.70 \\
& & & & & \\
\hline
\end{tabular}
\caption{Production cross sections (pb) in UED and SUSY models,
with UED or SUSY masses.
\label{tab:prod}}
}

Our numerical results for the production cross sections at the LHC are presented in
table~\ref{tab:prod}.  These results
were obtained from parton-level Monte Carlo simulations of the production processes
and decay chains, using the \HW\ event generator in SUSY mode with parton showering,
hadronization and underlying event switched off. The \HW\ default (MRST
leading-order~\cite{Martin:1998np}) parton distributions were used.
For the UED simulations, the SUSY
matrix element subroutine was replaced by a UED one and the SUSY particle data input
file consisted of UED data based on ref.~\cite{Cheng:2002iz}.

As a result of the more singular structure of the matrix elements and the extra helicity states,
the UED production cross sections tend to be larger than those of the analogous SUSY processes
for identical mass spectra, leading to an overall enhancement of the cross section, by
a factor of about 8 for both the mass scenarios that we studied.  Thus a SUSY-like
signature (e.g.\ many jets and leptons plus large missing energy) with a cross section
an order of magnitude larger than that predicted by SUSY models could be an
initial indication of UED.

The symbol $\sigma_{\rm all}$ represents the sum of cross section for all
two-to-two subprocesses included for that model, while $\sigma_{q^*}$ and
$\sigma_{\bar q^*}$ represent the inclusive cross sections for direct and
indirect production of the KK-quarks and -antiquarks (or squarks and antisquarks)
initiating the decay chains in figure 1.\footnote{Top-flavoured KK-quarks or
squarks were excluded as their decays would have a different signature.}
Although the overall magnitude of the cross sections is different in the UED
and SUSY models, we see that the KK-quark or squark fraction $f_q$ remains at
about 70\% for both models in both mass scenarios.

\section{Experimental observables}\label{sec:exp}

To investigate the observability of the effects discussed above, we switched on
the parton showering, hadronization and underlying event in our \HW\ simulations
and applied cuts to approximate those that might be used to select
new physics experimentally.  For jet cuts we used the simple calorimeter
simulation and cone jet finder GETJET~\cite{Paige_pc}, with cone size $\Delta R=0.7$.
Our cuts were as follows:
\begin{enumerate}
\item Missing transverse energy at least 100 GeV;
\item At least four jets with transverse energies ($E_T$) above 50 GeV;
\item Sum of missing $E_T$ and four highest jet  $E_T$'s at least 400 GeV;
\item Quark jet plus lepton invariant masses within the allowed range:
$m_{jl^\pm}\leq (m_{ql})_{\rm max}$.
\end{enumerate}
For cut 4 we cheated somewhat by selecting the jet that is nearest
(in $\Delta R$) to the true direction of the quark in the parton-level
decay chain, and the leptons that do belong to that chain.
Since we assume that all the new particle masses are already known, we expect that most
ambiguities in the reconstruction of decay chains would in fact be resolved by invariant
mass fits.

Of course more sophisticated cuts and detector simulations could be applied
to suit particular experiments, but this analysis should at least give an
indication of whether further efforts in that direction are warranted.

We included the UED spin correlations by generating decays according to phase space and
reweighting events according to the invariant expression (\ref{eq:uedspin}), evaluated
at the parton level. To improve efficiency, we forced the decay $Z^*\to l^*l$
($l=e$ or $\mu$), and correspondingly $\widetilde\chi_2^0\to\tilde l l$, thereby
enhancing the yield of the desired decay chains by factors of 3 and 8 in the UED
and SUSY mass scenarios, respectively.  To have comparable
samples of each type, we generated $2\times 10^5$ chains of each type in figure 1
(including their charge conjugates) for each mass scenario. However, because of the
different production cross sections and branching ratios in UED and SUSY, this
corresponds to varying effective luminosities, as summarised in table~\ref{tab:lumi}.

\TABLE{
\begin{tabular}{|c|c||c|c|c|c|c|}
\hline
Masses & Model & ${\cal L}_{\rm eff}$ & Cut 1 & 1+2 & 1+2+3 & 1+2+3+4 \\
\hline \hline
& & & & & & \\
UED    & UED  &   7 & 0.52 & 0.11 & 0.11 & 0.05\\
& & & & & & \\
UED    & SUSY &  66 & 0.53 & 0.13 & 0.12 & 0.06\\
& & & & & & \\
SPS 1a & UED  &  14 & 0.86 & 0.56 & 0.56 & 0.55\\
& & & & & & \\
SPS 1a & SUSY & 131 & 0.86 & 0.54 & 0.54 & 0.53\\
& & & & & & \\
\hline
\end{tabular}
\caption{Effective luminosities (fb$^{-1}$) of our sample of
$2\times 10^5$ decay chains, and fractions surviving the cuts listed above.
\label{tab:lumi}}
}

Also shown in table~\ref{tab:lumi} are the fractions of events with the desired decay
chains that survive the cuts listed above.  For the SUSY mass scenario, the effect of
these cuts is not great. However, in the UED mass scenario the near-degeneracy of the
mass spectrum means that the quark jet is likely to be relatively soft.  It is
therefore often misidentified or not found, with the result that few events survive
the jet cuts.

Figures \ref{fig:ued_mjl_det} and \ref{fig:sps_mjl_det} show the reconstructed jet plus
lepton mass distributions for the UED and SUSY mass scenarios, respectively. As we
saw in section~\ref{sec:uedspin}, the expected quasi-degeneracy of the UED mass
spectrum reduces the spin correlations and makes it hard to distinguish UED from
SUSY in that case, even at the parton level.  The experimental problems mentioned
above, due to the relatively soft quark jet, further reduce the sensitivity and
statistics.  Consequently the charge
asymmetry, shown in figure \ref{fig:asymm_det}(a), is unlikely to be observable in
the case of a UED-like mass spectrum.

\FIGURE{ \put(20,0){(a)}\epsfig{figure=ued_mjl+_det.ps,height=0.25\textheight}
  \put(20,0){(b)}\epsfig{figure=ued_mjl-_det.ps,height=0.25\textheight}
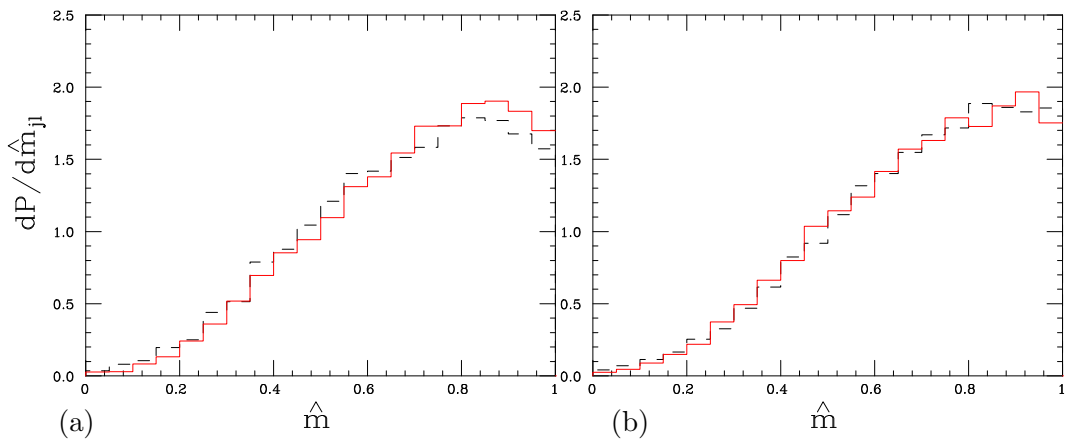
\caption{Detector-level rescaled mass distributions for (a) jet + $l^+$ (b) jet + $l^-$,
for the UED mass spectrum in table~\ref{tab:UEDmasses}. Dashed: SUSY. Solid/red: UED.
\label{fig:ued_mjl_det}}
}

\FIGURE{ \put(20,0){(a)}\epsfig{figure=sps_mjl+_det.ps,height=0.25\textheight}
  \put(20,0){(b)}\epsfig{figure=sps_mjl-_det.ps,height=0.25\textheight}
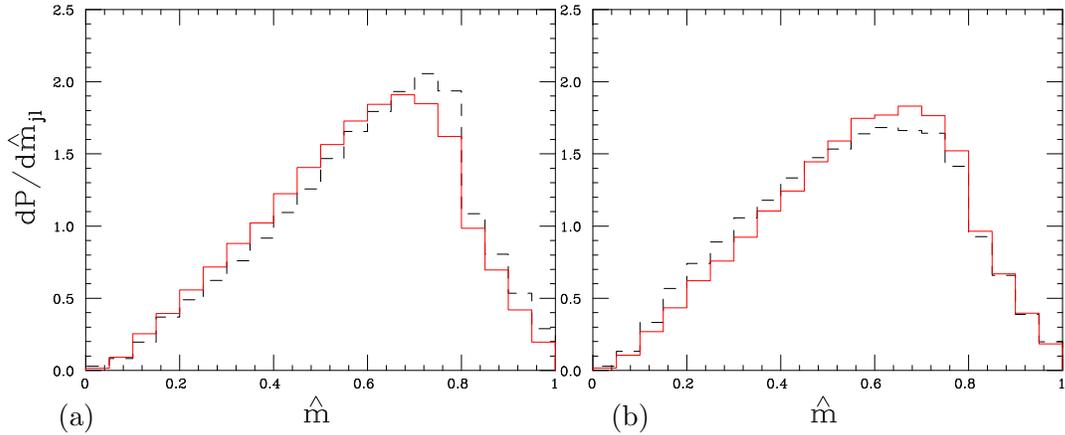
\caption{Detector-level rescaled mass distributions for (a) jet + $l^+$ (b) jet + $l^-$,
for the SUSY mass spectrum in table~\ref{tab:SUSYmasses}. Dashed: SUSY. Solid/red: UED.
\label{fig:sps_mjl_det}}
}

The results for the SUSY (SPS 1a) mass spectrum, figures \ref{fig:sps_mjl_det} and
\ref{fig:asymm_det}(b), are more encouraging. The spin correlations are larger in this
case and, apart from some resolution smearing,
their effects are not greatly diminished at the detector level.
Correspondingly the charge asymmetry remains visible and similar to that predicted at the
parton level, except at very high and low masses, where the asymmetry is the ratio of two
vanishing quantities.

\FIGURE{ \put(20,0){(a)} \epsfig{figure=ued_asym_det.ps,height=0.25\textheight}
  \put(20,0){(b)}\epsfig{figure=sps_asym_det.ps,height=0.25\textheight}
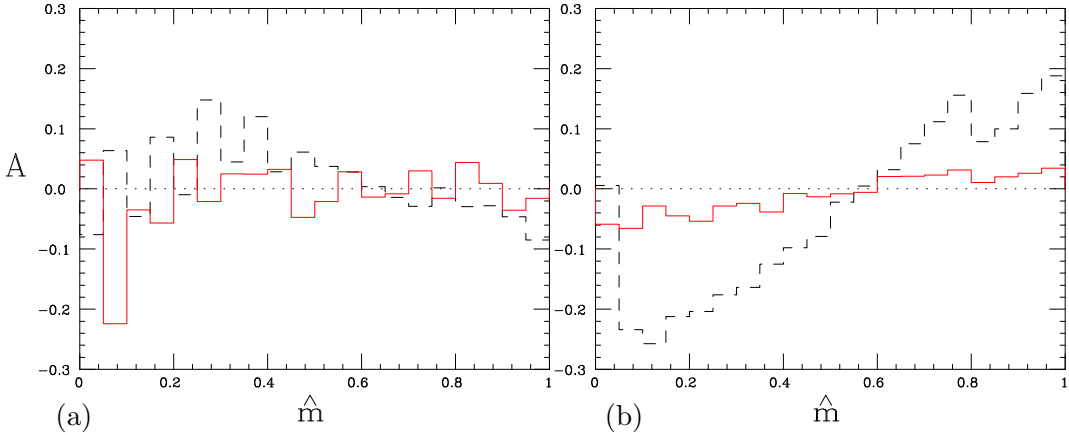
\caption{Detector-level charge asymmetries with respect to the jet + lepton rescaled invariant
mass, for the (a) UED and (b) SUSY mass spectra given above. Dashed: SUSY. Solid/red: UED.
\label{fig:asymm_det}}
}

%%%%%%%%%%%%%%%%%%%%%%%%%%%%%%%%%%%%%%%%%%%%%%%%%%%%%%%%%%%%%%%%%%%%%%
%%%%%%%%%%%%%%%%%%%%%%%%%% conclusions %%%%%%%%%%%%%%%%%%%%%%%%%%%%%%%
%%%%%%%%%%%%%%%%%%%%%%%%%%%%%%%%%%%%%%%%%%%%%%%%%%%%%%%%%%%%%%%%%%%%%%

\section{Conclusions}\label{sec:conc}

We have presented results of a comparative study of spin correlations
in models with supersymmetry and universal extra dimensions.  Complete 
results were obtained for a decay chain that is likely to be important if
either model is relevant at LHC energies.  The analytical expressions
for two-particle invariant mass distributions in section~\ref{sec:uedspin}
can be used to test the models for any combination of masses and chirality
of the new particles involved in the decay chain.  We presented numerical
and graphical results for two particular mass scenarios: one UED-like
and one SUSY-like (SPS 1a). In the former case the near-degeneracy of the
mass spectrum of new particles would make it difficult to verify the spin
content of the model in this way.  In SUSY models such degeneracy would
be less likely and the prospects for distinguishing between SUSY and
UED spin assignments are better.

As an adjunct to our study of spin effects we rederived the production
cross sections for KK-partons in UED models and found some differences
from results in the literature.  Due to the more singular matrix elements
and extra helicity states, the cross sections are significantly larger
than those for SUSY particles with the same mass spectrum. This could
also serve as a means of discriminating between SUSY and UED.

%%%%%%%%%%%%%%%%%%%%%%%%%%%%%%%%%%%%%%%%%%%%%%%%%%%%%%%%%%%%%%%%%%%%%%
%%%%%%%%%%%%%%%%%%%%%%%% acknowledgements %%%%%%%%%%%%%%%%%%%%%%%%%%%%
%%%%%%%%%%%%%%%%%%%%%%%%%%%%%%%%%%%%%%%%%%%%%%%%%%%%%%%%%%%%%%%%%%%%%%

\section*{Acknowledgements}

We thank colleagues in the Cambridge SUSY Working Group for helpful
discussions.  We are especially grateful to Chris Lester for assistance
in deriving the results on $ql^{\rm far}$ distributions in
appendix~\ref{app:qlfar}. We thank Cosmin Macesanu for valuable
communications which led to the correction of errors in an earlier
version of appendix~\ref{app:prod}. BRW thanks Mihoko Nojiri for stimulating
conversations and the Yukawa Institute, Kyoto University, and
the CERN Theory Group for hospitality while part of this work was
performed.

%%%%%%%%%%%%%%%%%%%%%%%%%%%%%%%%%%%%%%%%%%%%%%%%%%%%%%%%%%%%%%%%%%%%%%
%%%%%%%%%%%%%%%%%%%%%%%%%%% appendices %%%%%%%%%%%%%%%%%%%%%%%%%%%%%%%
%%%%%%%%%%%%%%%%%%%%%%%%%%%%%%%%%%%%%%%%%%%%%%%%%%%%%%%%%%%%%%%%%%%%%%

\section*{Appendices}
\appendix
\section{Quark + far lepton correlation}\label{app:qlfar}
\def\mqf{\mhat_{qf}}
The rescaled quark + far lepton invariant mass $\mqf$ is given in terms of the
decay angles by eq.~(\ref{eq:mqlfar}). The phase space distribution has the form
\beqn\label{eq:PS_qlf}
\frac{\ud P^{\rm PS}}{\ud \mqf} &=& -2\mqf\frac{\ln y}{1-y}
\qquad\;\;\mbox{for}\;0\leq \mqf\leq \sqrt{y}\nonumber\\
&=& -4\mqf\frac{\ln\mqf}{1-y}
\qquad\mbox{for}\; \sqrt{y} < \mqf\leq 1\;.
\eeqn

In the SUSY decay chain the only non-trivial dependence is on the angle $\theta^*$.
We find the following form for the rescaled mass distribution: for process 1
\beqn\label{eq:SUSY_qlf1}
\frac{\ud P^{\rm SUSY}_1}{\ud \mqf} &=& -\frac{4\mqf}{(1-y)^2}
(1-y+\ln y)\qquad\qquad\mbox{for}\;0\leq \mqf\leq \sqrt{y}\nonumber\\
&=& -\frac{4\mqf}{(1-y)^2}
(1-\mqf^2+2\ln\mqf)\;\;\;\mbox{for}\; \sqrt{y} < \mqf\leq 1
\eeqn
and for process 2
\beqn\label{eq:SUSY_qlf2}
\frac{\ud P^{\rm SUSY}_2}{\ud \mqf} &=& \frac{4\mqf}{(1-y)^2}
(1-y+y\ln y)\qquad\qquad\mbox{for}\;0\leq \mqf\leq \sqrt{y}\nonumber\\
&=& \frac{4\mqf}{(1-y)^2}
(1-\mqf^2+2y\ln\mqf)\;\;\;\;\mbox{for}\; \sqrt{y} < \mqf\leq 1\;.
\eeqn

In the case of UED, there is non-trivial dependence on all the decay angles.
We find the following form for the rescaled mass distribution: for process 1
\beqn\label{eq:UED_qlf1}
\frac{\ud P^{\rm UED}_1}{\ud \mqf} &=&  \frac{12\mqf}
{(1 + 2x)(2 + y)(1 + 2z)(1-y)^2}
\{(1 - y)[4x - y + 2(2 + 3y - 2x(5 + y))z \nonumber \\ &&- 
         4\mqf^2(2 - 3x)(1 - 2z)] - 
       [y(1 - 2(4 + y)z) +  4x(2z - y(1 - 4z)) \nonumber \\ &&+
         4\mqf^2(1 + y - x(2 + y))(1 - 2z)]\ln y\}
\qquad\mbox{for}\;0\leq \mqf\leq \sqrt{y}\nonumber\\
&=&  \frac{12\mqf}{(1 + 2x)(2 + y)(1 + 2z)(1-y)^2}
\{(1 - \mqf^2)(4x(1 + 2y - 5z - 6yz)-5y+2(2 + 9y)z \nonumber \\ && 
        - 4\mqf^2(1 - x)(1 - z)]  - 
      2[y(1 - 2(4 + y)z) +  4x(2z - y(1 - 4z))\nonumber\\&&
       +4\mqf^2(1 + y - x(2 + y))(1 - 2z)]\ln\mqf\}
\qquad\mbox{for}\; \sqrt{y} < \mqf\leq 1
\eeqn
and for process 2
\beqn\label{eq:UED_qlf2}
\frac{\ud P^{\rm UED}_2}{\ud \mqf} &=& \frac{12\mqf}
{(1 + 2x)(2 + y)(1 + 2z)(1-y)^2}
\{(1 - y)[-y + 2(2 + 2x(1 - y) + 3y)z \nonumber \\ &&- 
         4\mqf^2(2 - x)(1 - 2z)] -[ y(1 - 2(4 + y)z)\nonumber \\ &&
   +4\mqf^2(1 + (1 - x)y)(1 - 2z)]\ln y\}
\qquad\mbox{for}\;0\leq \mqf\leq \sqrt{y}\nonumber\\
&=&  \frac{12\mqf}{(1 + 2x)(2 + y)(1 + 2z)(1-y)^2}
\{(1 - \mqf^2)[4(1 + x)z - y(5 - 18z + 8xz)\nonumber \\ &&-
         4\mqf^2(1 - x)(1 - z)] - 2[y(1 - 2(4 + y)z)\nonumber \\ &&+
         4\mqf^2(1 + (1 - x)y)(1 - 2z)]\ln\mqf\}
\qquad\mbox{for}\; \sqrt{y} < \mqf\leq 1\;.
\eeqn

\section{UED production cross sections}\label{app:prod}
We neglect all Standard Model particle masses and work at tree level,
so that the $n^{th}$ excited KK state of each particle has mass $n/R$.
In practice we ignore $n>1$ excitations. We use here the notation of
ref.~\cite{Macesanu:2002db}:
$q_1^{\bullet}$ or $q_1^{\circ}$
represent the first KK-excitations of the 5D fields whose zero-modes are the
left-handed doublet and right-handed singlet quarks respectively. 
Throughout, $q_1^*$ represents either $q_1^{\bullet}$ or $q_1^{\circ}$,
$M_1=1/R$, $s,t,u$ are the usual Mandelstam variables, $t_3=t-M_1^2$ and $u_4=u-M_1^2$.
Only (\ref{qqb2qpqbp}), (\ref{qqb2qqb}) and (\ref{qqbp2qqbp}) agree in form
with ref.~\cite{Macesanu:2002db}.
In addition we find an extra factor of 16 in the overall normalisation.

\begin{equation}
\overline{\sum}|\mathcal{M}(q \bar{q} \to q_1^{* \prime} \bar{q}_1^{* \prime})|^2=\frac{4 g_s^4}{9} \left[ \frac{2 M_1^2}{s} + \frac{t_3^2+u_4^2}{s^2} \right], \label{qqb2qpqbp}
\end{equation}

\begin{eqnarray}
\nonumber \overline{\sum}|\mathcal{M}(q\bar{q} \to q_1^{*}\bar{q}_1^*)|^2&=&\frac{g_s^4}{9}\left[ 2 M_1^2 \left( \frac{4}{s}+\frac{s}{t_3^2}-\frac{1}{t_3} \right) \right. \\ && \left. + \frac{23}{6} + \frac{2s^2}{t_3^2}+\frac{8s}{3 t_3} + \frac{6 t_3}{s} + \frac{8 t_3^2}{s^2} \right],\label{qqb2qqb}
\end{eqnarray}

\begin{eqnarray}
\nonumber \overline{\sum}|\mathcal{M}(qq \to q_1^{*} q_1^*)|^2&=&\frac{g_s^4}{27} \left[ M_1^2 \left( 6\frac{t_3}{u_4^2}+6\frac{u_4}{t_3^2}-\frac{s}{t_3u_4}\right)\right. \\ && \quad \left. +2\left( 3\frac{t_3^2}{u_4^2}+3\frac{u_4^2}{t_3^2} + 4\frac{s^2}{t_3 u_4}-5 \right) \right],\label{qq2qq}
\end{eqnarray}

\begin{eqnarray}
\nonumber \overline{\sum}|\mathcal{M}(gg \to q_1^{*}\bar{q}_1^*)|^2&=&g_s^4 \left[M_1^4 \frac{-4}{t_3u_4} \left( \frac{s^2}{6 t_3 u_4}-\frac{3}{8} \right) + M_1^2 \frac{4}{s} \left(\frac{s^2}{6t_3 u_4}-\frac{3}{8} \right) \right. \\ && \qquad \quad \left. + \frac{s^2}{6 t_3 u_4} -\frac{17}{24}+\frac{3 t_3 u_4}{4s^2}\right], \label{gg2qqb}
\end{eqnarray}

\begin{equation}
\overline{\sum}|\mathcal{M}(gq \to g^* q_1^{*})|^2=\frac{-g_s^4}{3} \left[ \frac{5s^2}{12 t_3^2}+\frac{s^3}{t_3^2 u_4}+\frac{11 s u_4}{6 t_3^2}+\frac{5u_4^2}{12 t_3^2}+\frac{u_4^3}{s t_3^2}\right],\label{qg2qg}
\end{equation}

\begin{equation}
\overline{\sum}|\mathcal{M}(q \bar{q}^{\prime} \to q_1^{*} \bar{q}_1^{* \prime}  )|^2=\frac{g_s^4}{18} \left[ 4 M_1^2 \frac{s}{t_3^2}+5+4\frac{s^2}{t_3^2}+8\frac{s}{t_3} \right],\label{qqbp2qqbp}
\end{equation}

\begin{equation}
\overline{\sum}|\mathcal{M}(qq^{\prime} \to q_1^{*} q_1^{* \prime})|^2=\frac{2 g_s^4}{9} \left[ - M_1^2 \frac{s}{t_3^2}+\frac{1}{4}+\frac{s^2}{t_3^2} \right],\label{qqp2qqp}
\end{equation}

\begin{equation}
\overline{\sum}|\mathcal{M}(qq \to q_1^{\bullet} q_1^{\circ})|^2=\frac{g_s^4}{9}\left[ M_1^2 \left( \frac{2s^3}{t_3^2 u_4^2} - \frac{4s}{t_3 u_4} \right)+ 2\frac{s^4}{t_3^2 u_4^2} - 8\frac{s^2}{t_3 u_4}+5 \right],\label{qq2qbullqcirc}
\end{equation}

\begin{equation}
\overline{\sum}|\mathcal{M}(q \bar{q}^{\prime} \to q_1^{\bullet} \bar{q}_1^{\prime \circ} )|^2=\frac{g_s^4}{9} \left[ 2 M_1^2 \left( \frac{1}{t_3}+\frac{u_4}{t_3^2} \right) + \frac{5}{2} + \frac{4 u_4}{t_3}+\frac{2 u_4^2}{t_3^2} \right],\label{qqb2qbullqbcirc}
\end{equation}
which is the same result as for $q \bar{q} \to q_1^{\bullet}\bar{q}_1^{\circ},q_1^{\circ}\bar{q}_1^{\bullet}$ and $q \bar{q}^{\prime} \to q_1^{\circ} \bar{q}_1^{\prime \bullet}$ as these are given by the same $t$-channel diagram.

\begin{equation}
\overline{\sum}|\mathcal{M}(q q^{\prime} \to q_1^{\bullet} q_1^{\prime \circ} )|^2=\frac{g_s^4}{9} \left[ -2 M_1^2 \left( \frac{1}{t_3}+\frac{u_4}{t_3^2} \right) + \frac{1}{2} +\frac{2 u_4^2}{t_3^2} \right],\label{qq2qcircqbull}
\end{equation}
which is the same result as for $q q^{\prime} \to q_1^{\circ} q_1^{\prime
  \bullet}$.\footnote{We are grateful to Martyn Gigg and Peter Richardson for pointing out that the result for these processes had mistakenly been
 taken to be the same as equation
  (\ref{qqb2qbullqbcirc}) in an earlier version.}

\begin{eqnarray}
\nonumber \overline{\sum}|\mathcal{M}(g g \to g^* g^*)|^2&=&\frac{9 g_s^4}{4} \left[ 3 M_1^4 \frac{s^2+t_3^2+u_4^2}{t_3^2 u_4^2}- 3 M_1^2 \frac{s^2+t_3^2+u_4^2}{s t_3 u_4} \right. \\ &&\qquad \quad \left. +1+\frac{(s^2+t_3^2+u_4^2)^3}{4s^2t_3^2u_4^2}-\frac{t_3 u_4}{s^2} \right],\label{gg2gg}
\end{eqnarray}

\begin{eqnarray}
\nonumber \overline{\sum}|\mathcal{M}(q \bar{q} \to g^* g^*)|^2&=&\frac{2 g_s^4}{27} \left[M_1^2 \left( -\frac{4s^3}{t_3^2 u_4^2}+\frac{57s}{t_3u_4}-\frac{108}{s} \right) \right. \\ && \qquad \quad \left. + \frac{20s^2}{t_3u_4}-93+\frac{108t_3 u_4}{s^2} \right].\label{qqb2gg}
\end{eqnarray}

%%%%%%%%%%%%%%%%%%%%%%%%%%%%%%%%%%%%%%%%%%%%%%%%%%%%%%%%%%%%%%%%%%%%%%
%%%%%%%%%%%%%%%%%%%%%%%%% bibliography %%%%%%%%%%%%%%%%%%%%%%%%%%%%%%%
%%%%%%%%%%%%%%%%%%%%%%%%%%%%%%%%%%%%%%%%%%%%%%%%%%%%%%%%%%%%%%%%%%%%%%


\begin{thebibliography}{100}

%\cite{Barr:2004ze}
\bibitem{Barr:2004ze}
A.~J.~Barr,
{\it Determining the spin of supersymmetric particles at the LHC using lepton charge asymmetry},
Phys.\ Lett.\ B {\bf 596} (2004) 205
[arXiv:hep-ph/0405052].
%%CITATION = HEP-PH 0405052;%%

%\cite{Allanach:2000kt}
\bibitem{Allanach:2000kt}
B.~C.~Allanach, C.~G.~Lester, M.~A.~Parker and B.~R.~Webber,
{\it Measuring sparticle masses in non-universal string inspired models at  the LHC},
JHEP {\bf 0009} (2000) 004
[arXiv:hep-ph/0007009].
%%CITATION = HEP-PH 0007009;%%

%\cite{Lester:2001zx}
\bibitem{Lester:2001zx}
C.~G.~Lester,
{\it Model independent sparticle mass measurements at ATLAS},
CERN-THESIS-2004-003
%\href{http://www.slac.stanford.edu/spires/find/hep/www?r=cern-thesis-2004-003}{SPIRES entry}

%\cite{Appelquist:2000nn}
\bibitem{Appelquist:2000nn}
T.~Appelquist, H.~C.~Cheng and B.~A.~Dobrescu,
{\it Bounds on universal extra dimensions},
Phys.\ Rev.\ D {\bf 64} (2001) 035002
[arXiv:hep-ph/0012100].
%%CITATION = HEP-PH 0012100;%%

%\cite{Cheng:2002iz}
\bibitem{Cheng:2002iz}
H.~C.~Cheng, K.~T.~Matchev and M.~Schmaltz,
{\it Radiative corrections to Kaluza-Klein masses},
Phys.\ Rev.\ D {\bf 66} (2002) 036005
[arXiv:hep-ph/0204342];
%%CITATION = HEP-PH 0204342;%%
%\cite{Cheng:2002ab}
%\bibitem{Cheng:2002ab}
%H.~C.~Cheng, K.~T.~Matchev and M.~Schmaltz,
{\it Bosonic supersymmetry? Getting fooled at the LHC},
\ibid {\bf 66} (2002) 056006
[arXiv:hep-ph/0205314].
%%CITATION = HEP-PH 0205314;%%

%\cite{Goto:2004cp}
\bibitem{Goto:2004cp}
T.~Goto, K.~Kawagoe and M.~M.~Nojiri,
{\it Study of the slepton non-universality at the CERN Large Hadron Collider},
Phys.\ Rev.\ D {\bf 70}, 075016 (2004)
[Erratum-ibid.\ D {\bf 71}, 059902 (2005)]
[arXiv:hep-ph/0406317].
%%CITATION = HEP-PH 0406317;%%

%\cite{Bhattacharyya:2005vm}
\bibitem{Bhattacharyya:2005vm}
G.~Bhattacharyya, P.~Dey, A.~Kundu and A.~Raychaudhuri,
{\it Probing universal extra dimension at the International Linear Collider},
arXiv:hep-ph/0502031.
%%CITATION = HEP-PH 0502031;%%

%\cite{Battaglia:2005zf}
\bibitem{Battaglia:2005zf}
M.~Battaglia, A.~Datta, A.~De Roeck, K.~Kong and K.~T.~Matchev,
{\it Contrasting supersymmetry and universal extra dimensions at the CLIC multi-TeV e+ e- collider},
arXiv:hep-ph/0502041.
%%CITATION = HEP-PH 0502041;%%

%\cite{Macesanu:2002db}
\bibitem{Macesanu:2002db}
C.~Macesanu, C.~D.~McMullen and S.~Nandi,
{\it Collider implications of universal extra dimensions},
Phys.\ Rev.\ D {\bf 66} (2002) 015009
[arXiv:hep-ph/0201300].
%%CITATION = HEP-PH 0201300;%%

%\cite{Corcella:2000bw}
\bibitem{Corcella:2000bw}
G.~Corcella, I.G.~Knowles, G.~Marchesini, S.~Moretti, 
  K.~Odagiri, P.~Richardson, M.~H.~Seymour and B.~R.~Webber,
{\it HERWIG 6: An event generator for hadron emission reactions with  interfering gluons (including supersymmetric processes)}, JHEP {\bf 0101} (2001) 010
[arXiv:hep-ph/0011363].
%%CITATION = HEP-PH 0011363;%%


%\cite{Corcella:2002jc}
\bibitem{Corcella:2002jc}
G.~Corcella {\it et al.},
{\it HERWIG 6.5 release note},
arXiv:hep-ph/0210213.
%%CITATION = HEP-PH 0210213;%%
%\cite{Barr:2004ze}

%\cite{Richardson:2001df}
\bibitem{Richardson:2001df}
P.~Richardson,
{\it Spin correlations in Monte Carlo simulations},
JHEP {\bf 0111} (2001) 029
[arXiv:hep-ph/0110108].
%%CITATION = HEP-PH 0110108;%%

%\cite{Moretti:2002eu}
\bibitem{Moretti:2002eu}
S.~Moretti, K.~Odagiri, P.~Richardson, M.~H.~Seymour and B.~R.~Webber,
{\it Implementation of supersymmetric processes in the HERWIG event  generator},
JHEP {\bf 0204}, 028 (2002)
[arXiv:hep-ph/0204123].
%%CITATION = HEP-PH 0204123;%%

%\cite{Allanach:2002nj}
\bibitem{Allanach:2002nj}
B.~C.~Allanach {\it et al.},
{\it The Snowmass points and slopes: Benchmarks for SUSY searches},
in {\it Proc. of the APS/DPF/DPB Summer Study on the Future of Particle Physics (Snowmass 2001) } ed. N.~Graf,
Eur.\ Phys.\ J.\ C {\bf 25} (2002) 113
[eConf {\bf C010630} (2001) P125]
[arXiv:hep-ph/0202233].
%%CITATION = HEP-PH 0202233;%%

%\cite{Pukhov:1999gg}
\bibitem{Pukhov:1999gg}
A.~Pukhov {\it et al.},
{\it CompHEP: A package for evaluation of Feynman diagrams and integration  over
multi-particle phase space. User's manual for version 33},
arXiv:hep-ph/9908288.
%%CITATION = HEP-PH 9908288;%%

%\cite{Martin:1998np}
\bibitem{Martin:1998np}
A.~D.~Martin, R.~G.~Roberts, W.~J.~Stirling and R.~S.~Thorne,
{\it Scheme dependence, leading order and higher twist studies of MRST  partons},
Phys.\ Lett.\ B {\bf 443} (1998) 301
[arXiv:hep-ph/9808371].
%%CITATION = HEP-PH 9808371;%%

\bibitem{Paige_pc}
F.~Paige, private communication.
\end{thebibliography}
\end{document}